# Physical properties of predicted Ti$_2$CdN versus existing Ti$_2$CdC MAX phase: An *ab initio* study

M. Roknuzzaman[1], M. A. Hadi[2], M. J. Abden[3], M. T. Nasir[2], A. K. M. A. Islam[2,4],
M. S. Ali[5], K. Ostrikov[6], and S. H. Naqib[2]*

[1]Department of Physics, Jessore University of Science and Technology, Jessore 7408, Bangladesh
[2]Department of Physics, Rajshahi University, Rajshahi 6205, Bangladesh
[3]Department of EEE, International Islamic University Chittagong, Chittagong 4203, Bangladesh
[4]International Islamic University Chittagong, Chittagong 4203, Bangladesh
[5]Department of Physics, Pabna University of Science and Technology, Pabna 6600, Bangladesh
[6]School of Chemistry, Physics and Mechanical Engineering, Queensland University of Technology, QLD 4000, Australia

**Abstract**

*Ab intio* calculations were done to investigate the structural, elastic, electronic and optical properties of the Cd-containing *theoretically* predicted MAX phase, Ti$_2$CdN, in comparison with the isostructural and already synthesized phase, Ti$_2$CdC. These calculations reveal that the substitution of C by N affects the lattice parameter *c*, whereas the lattice parameter *a*, remains almost unchanged. All the elastic constants and moduli increase when carbon is replaced by nitrogen. The elastic anisotropy in Ti$_2$CdC is higher in comparison with that of Ti$_2$CdN. Both these nanolaminates are brittle in nature. The calculated electronic band structures and density of states suggest that the chemical bonding in these two ternary compounds is a combination of covalent, ionic and metallic in nature. Electrical conductivity of Ti$_2$CdC is found to be higher than that of Ti$_2$CdN. The calculated reflectivity spectra show that both the MAX phases Ti$_2$CdC and Ti$_2$CdN have the potential to be used as coating materials to minimize solar heating.

## 1. Introduction

The MAX phases, a class of layered ternary carbides and nitrides, are denoted by the chemical formula $M_{n+1}AX_n$, having different MAX stoichiometries. These are generally referred to as 211, 312, and 413 phases for n = 1, 2, and 3, respectively. In the periodic table of elements, M is an early transition metal from groups 3 (Sc), 4 (Ti, Zr, Hf), 5 (V, Nb, Ta), or 6 (Cr, Mo), A is an A-group element from groups 12 (Cd), 13 (Al, Ga, In, Tl), 14 (Si, Ge, Sn, Pb), 15 (P, As), or 16 (S), and X is C and/or N. MAX compounds are thermodynamically stable nanolaminates which have significant potential for industrial applications owing to their remarkable combination of physical, chemical, electrical, and mechanical properties, some of which are characteristics of metals and some of ceramics [1]. The MAX phases serve as effective nanolaminate materials because of their laminated layers having thickness in the nanometer range. Due to their simultaneous metallic and ceramics characteristics, the MAX phases are often termed as metallic ceramics [2]. These technologically important compounds have drawn major attention of materials scientists, physicists, and chemists because of their good electrical and thermal conductivities, damage tolerance, machinability, and thermal shock resistance. In addition, these materials show high resistance to oxidation and corrosion, good elastic rigidity, and ability to maintain the strengths at high temperature [2-9]. The MAX phases are developed for defense, aerospace, automotive applications, medical, and portable electronic devices. A detailed study of such materials is important as the MAX phases have many advantages over the commonly used materials. A recent review on MAX phase nanolaminates can be found in Ref. [10].

So far more than 70 synthesized MAX compounds exist [11]. Among those only Ti$_2$CdC phase contains Cd as A-group element. Moreover, this material was synthesized at the early period of discovery of the MAX phases [12]. In spite of that, Ti$_2$CdC is a relatively less studied member of 211 phases. Liu et al. [13] carried out a theoretical study on structural and mechanical properties of Ti$_2$CdC together with other Ti$_2$AC (A = Sn, Ga, In, and Pb) phases. An *ab initio* study, by Bai et al. [14], reported on the chemical bonding by means of electronic density of states (DOS) calculations. They also calculated the elastic properties and the structural parameters of Ti$_2$CdC. An earlier theoretical study predicted Ti$_2$CdN as a stable nanolaminate [15] from the calculated values of the elastic constants. However, full potential of this material still remains poorly understood. Therefore, further theoretical study is needed to stimulate experimental research on its synthesis and applications. This motivates us to perform the present study on the structural, elastic, electronic, and optical properties of these Cd-containing MAX phases. These studies are expected to shed light on the effect of substitution of C by N in Ti$_2$CdC phase, in particular.

---
* Corresponding author: salehnaqib@yahoo.com

At this point, it should be mentioned that in recent years awareness regarding biohazard is increasing. Cd containing compounds are considered as toxic. On the other hand, a large number of technologically important devices (e.g., battery, solar cells, and other semiconducting compounds and thin films for various device applications) contain Cd. Both the MAX phases under study contain Cd and therefore, should preferably be used under controlled environment.

The rest of this paper is divided into three sections. In section 2, a brief description of the computation procedures used in this investigation have been presented. The results obtained for the structural, elastic, electronic, and optical properties of Ti$_2$CdC and Ti$_2$CdN are analyzed in section 3. Finally, section 4 summarizes the main conclusions of the present work.

## 2. Computation procedures

The first-principles investigations are conducted by using the Cambridge Serial Total Energy Package (CASTEP) code [16], in which the plane-wave pseudopotential total energy calculation approach based on the density functional theory (DFT) [17, 18] is implemented. The electronic exchange-correlation energy is treated according to the Perdew-Burke-Ernzerhof generalized gradient approximation (PBE-GGA) [19]. Interactions of electrons with ion cores are taken into account by using the Vanderbilt-type ultrasoft pseudopotential [20]. A plane-wave cutoff energy of 500 eV is used throughout the calculations. The crystal structures are fully optimized to determine the various ground state properties. The geometry optimizations are performed through minimizing the total energy and internal forces by using the Broyden-Fletcher-Goldfarb-Shanno (BFGS) minimization technique [21]. The convergence criteria are selected as the difference in total energy being less than 5 × 10$^{-6}$ eV/atom, the maximum ionic Hellmann-Feynman force being less than 0.01 eV/Å, maximum ionic displacement being less than 5 × 10$^{-4}$ Å, and the maximum stress being less than 0.02 GPa. The energy tolerance in the self-consistent field calculation is set to 5 × 10$^{-7}$ eV/atom. For the sampling of the Brillouin zone, the widely used Monkhorst-Pack scheme [22] is employed to produce a uniform grid of points in the k-space along the three axes in reciprocal space. 15 × 15 × 3 k-points are taken for both Ti$_2$CdC and Ti$_2$CdN to achieve the ground-states. In optical properties calculations, the Vanderbilt-type ultrasoft pseudopotential is replaced by the norm-conserving pseudopotential, keeping the other setup completely unchanged.

The elastic constants are obtained by applying a set of finite homogeneous deformations and calculating the resulting stresses [23], as implemented within the CASTEP code.

The optical properties of a medium at all frequencies can be described entirely by the dielectric function, $\varepsilon(\omega) = \varepsilon_1(\omega) + i\varepsilon_2(\omega)$, which is related to the interactions of photons with electrons. The imaginary part, $\varepsilon_2(\omega)$, of the dielectric function is calculated in the momentum representation of matrix elements between occupied and unoccupied electronic eigenstates by using the CASTEP supported formula given by

$$\varepsilon_2(\omega) = \frac{2e^2\pi}{\Omega\varepsilon_0} \sum_{k,v,c} \left|\left\langle \psi_k^c | \hat{u}\cdot\vec{r} | \psi_k^v \right\rangle\right|^2 \delta(E_k^c - E_k^v - E), \qquad (1)$$

where $\Omega$ is the unit cell volume, $\omega$ is the frequency of light, $e$ is the charge of an electron, $\hat{u}$ is the unit vector defining the polarization of the incident electric field and $\psi_k^c$ and $\psi_k^v$ are the conduction and valence band wave functions at a given $k$, respectively. The Kramers-Kronig transformations give the real part $\varepsilon_1(\omega)$ of the dielectric function from the imaginary part. The refractive index, the absorption spectrum, the energy loss-function, the reflectivity, and the optical conductivity (the real part) are derived once $\varepsilon_1(\omega)$ and $\varepsilon_2(\omega)$ are known [24]. The intraband contribution to the optical properties of metallic compounds like the MAX phases under consideration, affects mainly the low-energy part of the spectra. It can be corrected for the dielectric function by including an empirical Drude term [25, 26] with unscreened plasma frequency of 3.0 eV and damping of 0.05 eV.

## 3. Results and discussion

### 3.1. *Structural properties*

The MAX compounds $Ti_2CdC$ and $Ti_2CdN$ belong to the hexagonal crystallographic system with space group $P6_3/mmc$ (No. 194). There are two formula units and eight atoms per unit cell (Fig. 1). Fully relaxed structures for both these layered nanolaminates are obtained by optimizing the geometry including the lattice constants and internal atomic positions. The optimized Ti atom is located on the $4f$ Wyckoff position with fractional coordinates (1/3, 2/3, 0.077) and (1/3, 2/3, 0.076) for $Ti_2CdC$ and $Ti_2CdN$, respectively. The Cd atoms are positioned in the $2d$ Wyckoff site with fractional coordinates (1/3, 2/3, 3/4) in $Ti_2CdC$ and $Ti_2CdN$. The C/N atoms are situated in the $2a$ Wyckoff position with fractional coordinates (0, 0, 0). The lattice constants $a$ and $c$, internal atomic coordinate $z$, equilibrium unit cell volume $V$, Bulk modulus $B$, and pressure derivative of bulk modulus $B'$ for $Ti_2CdC$ and $Ti_2CdN$ at 0 K are given in Table 1. The structural parameters calculated in this study agree quite well with both experimental (where applicable) and theoretical values. As can be noted from Table 1, the substitution of C by N mostly affects the $c$ values; the $a$ values remain almost unchanged. This is opposite to the general trend (as C is replaced by N, lattice parameter $a$ changes more in comparison with the $c$ values [27]). To understand this intriguing discrepancy, more theoretical work is required. The bulk modulus increases by 12.5% when C is substituted by N.

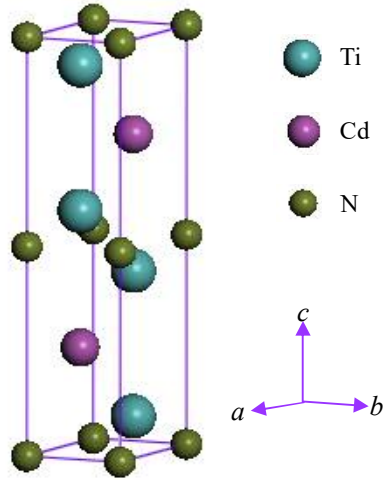

**Fig. 1.** The unit cell of $Ti_2CdN$ as a structural model of 211 MAX phases.

**Table 1.** Calculated lattice parameters *a* and *c* (in Å), hexagonal ratio *c/a*, internal parameter *z*, unit cell volume *V* (in Å$^3$), bulk modulus *B* (in GPa) and its pressure derivative *B′* for Ti$_2$CdC and Ti$_2$CdN.

| Compound | *a* | *c* | *c/a* | *z* | *V* | *B* | *B′* | Remarks |
|---|---|---|---|---|---|---|---|---|
| Ti$_2$CdC | 3.103 | 14.57 | 4.696 | 0.077 | 121.5 | 114 | 4.9 | This study |
|  | 3.099 | 14.41 | 4.650 | 0.086 | 119.8 | - | - | Expt. [12] |
|  | 3.091 | 14.528 | 4.700 | - | 120.2 | - | - | Calc. [13] |
|  | 3.104 | 14.55 | 4.687 | 0.077 | 121.4 | 116 | - | Calc. [14] |
|  | 3.106 | 14.54 | 4.681 | - | 121.5 | 116 | - | Calc. [15] |
| Ti$_2$CdN | 3.078 | 14.19 | 4.610 | 0.076 | 116.5 | 127 | 5.4 | This study |
|  | 3.082 | 14.15 | 4.591 | - | 116.4 | 130 | - | Calc. [15] |

### 3.2. *Elastic properties*

Table 2 shows the calculated elastic constants $C_{ij}$, together with other existing theoretical results for Ti$_2$CdC and Ti$_2$CdN. The calculated results for both the nanolaminates agree well with the values found in literatures [12-15]. The elastic constants are positive and satisfy the well-established Born criteria [28], which suggests that the two 211 MAX phases under study are mechanically stable. It is observed that the elastic constants $C_{11}$ and $C_{33}$ increase when C is substituted with N. Thus, one can conclude that the Ti-N bonds are stronger than the Ti-C bonds. The calculated shear anisotropy factor, *A*, defined as $A = 4C_{44}/(C_{11} + C_{33} - 2C_{13})$, implies that Ti$_2$CdC possesses larger anisotropy compared to Ti$_2$CdN for the shear planes $\{1\,0\,\bar{1}\,0\}$, between the directions $\langle 0\,1\,\bar{1}\,1 \rangle$ and $\langle 0\,1\,\bar{1}\,0 \rangle$.

Anisotropy parameter can also be defined by the ratio between the linear compressibility coefficients along the c- and a-axis for hexagonal crystal: $k_c/k_a = (C_{11} + C_{12} - 2C_{13})/(C_{33} - C_{13})$. Calculated values show that in comparison with Ti$_2$CdN, Ti$_2$CdC is more compressible along the c-axis. In fact, the elastic anisotropy for Ti$_2$CdN is quite mild and this phase is almost cubic [27].

Table 2 also lists the values of bulk modulus *B*, shear modulus *G*, Young modulus *Y* and Poisson's ratio *v* of the two compounds. Bulk and shear moduli are the most important parameters for estimating the material hardness. The small bulk and shear moduli of Ti$_2$CdC and Ti$_2$CdN imply that they can be grouped into high compressibility materials. Young's modulus is used to measure the stiffness of a solid. Theoretical results show Ti$_2$CdN (*Y* = 212 GPa) is significantly stiffer than Ti$_2$CdC (*Y* = 162 GPa).

Solids can be classified into two groups according to the Pugh's criteria [29]: ductile and brittle. A material is classified as brittle if its *G/B* > 0.5, otherwise it should be ductile. The Poisson's ratio for a ductile metallic material is typically 0.33 and for a brittle material, it is less than 0.33 [30]. Thus according to Pugh's criteria and the values of Poisson's ratio, both Ti$_2$CdC and Ti$_2$CdN are brittle in nature which is the general behavior of MAX phases [31-33]. The relatively low values of the Poisson's ratios for the two compounds indicate their high degree of directional covalent bonding. The directional covalent bonding tends to increase shear modulus.

The phase stability of various MAX phases is an important issue. The Born criteria mainly concern itself with mechanical stability. The possibility of thermodynamically stable phase formation and the tendency of decomposition into competing phases are better studied via the enthalpy (*H*) of formation [34]. The calculated enthalpies for already synthesized Ti$_2$CdC and theoretically predicted Ti$_2$CdN are very close, -2036.64 KJ/mole and -2069.48 KJ/mole, respectively. Considering the binary compounds TiC, TiN, and TiCd as competing phases, the difference in the enthalpy of formation for Ti$_2$CdC, given by $\Delta H = H(\text{Ti}_2\text{CdC}) - H(\text{TiCd}) - H(\text{TiC})$, is found to be -62.82 KJ/mole. $\Delta H$ for Ti$_2$CdN is -63.25 KJ/mole. This analysis shows that the predicted phase should be as stable as the already synthesized phase.

**Table 2.** Calculated elastic constants $C_{ij}$ (GPa), bulk modulus $B$ (GPa), shear modulus $G$ (GPa), Young's modulus $Y$ (GPa), Poisson's ratio $v$, elastic anisotropic factor $A$ and $k_c/k_a$ of Ti$_2$CdC and Ti$_2$CdN.

| Compound | $C_{11}$ | $C_{12}$ | $C_{13}$ | $C_{33}$ | $C_{44}$ | $B$ | $G$ | $Y$ | $G/B$ | $v$ | $A$ | $k_c/k_a$ | Remarks. |
|---|---|---|---|---|---|---|---|---|---|---|---|---|---|
| Ti$_2$CdC | 257 | 68 | 44 | 205 | 36 | 114 | 64 | 162 | 0.56 | 0.26 | 0.385 | 1.472 | This study |
|  | 263 | 64 | 45 | 212 | 39 | 114 | 74 | 183 | 0.65 | 0.23 | 0.405 | 1.419 | Calc.[13] |
|  | 258 | 68 | 46 | 205 | 33 | 116 | 70 | 174 | 0.60 | 0.25 | 0.356 | 1.465 | Calc.[14] |
|  | 253 | 71 | 47 | 203 | 31 | 116 | 67 | 168 | 0.58 | 0.26 | 0.343 | 1.474 | Calc.[15] |
| Ti$_2$CdN | 266 | 76 | 56 | 235 | 74 | 127 | 87 | 212 | 0.68 | 0.22 | 0.761 | 1.285 | This study |
|  | 270 | 83 | 58 | 235 | 68 | 130 | 84 | 208 | 0.65 | 0.23 | 0.699 | 1.339 | Calc.[15] |

*3.3. Electronic properties*

The results of band structure calculations for Ti$_2$CdC and Ti$_2$CdN along the high symmetry directions in the first Brillouin zone are presented in Fig. 2. The Fermi level of both the nanolaminates lies below the valence band maximum near the $\Gamma$ point. The occupied valence bands of Ti$_2$CdC lie inside the energy range from -6.9 eV to Fermi level, $E_F$. On the other hand, in Ti$_2$CdN the valence bands extend from -7.2 eV to $E_F$. Moreover, for both the phases, a lot of valence bands go across the Fermi level and overlap with the conduction bands. Consequently, no band gap is found at the Fermi level and Ti$_2$CdC as well as Ti$_2$CdN (is predicted to) show metallic conductivity.

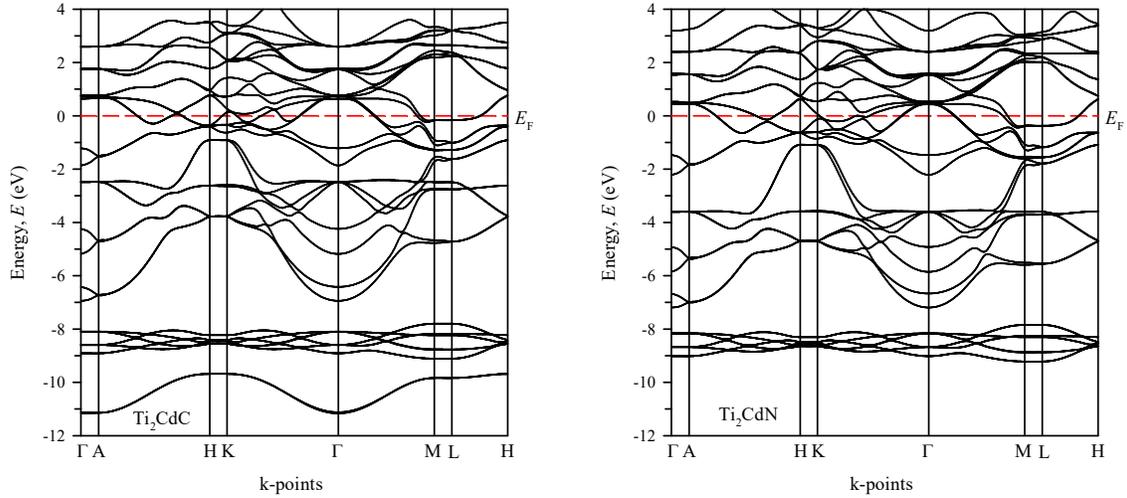

**Fig. 2.** Electronic band structure of (a) Ti$_2$CdC and (b) Ti$_2$CdN.

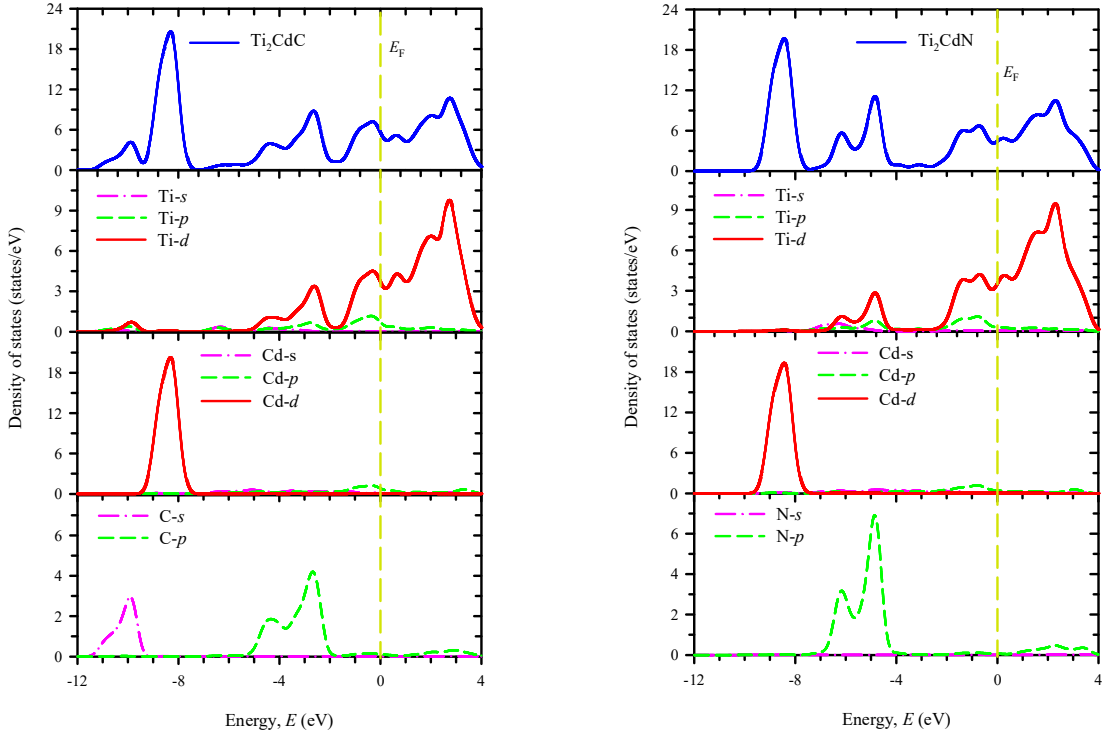

**Fig. 3**. Total and partial electronic density of states of (a) Ti$_2$CdC and (b) Ti$_2$CdN.

The total and partial electronic density of states (DOS) of Ti$_2$CdX (X = C or N) are presented in Fig. 3. It can be seen that the lowest-lying bands in Ti$_2$CdC, from -11.5 to -9.4 eV, are formed by the C 2s states with a small mixture of the Ti 3p and Ti 3d states. The highest valence bands occupied the energy range from -9.4 to -7.2 eV are derived almost entirely from Cd 4d states. The valence bands located between -7.2 and -5.4 eV below the Fermi level arise mainly from mixed Ti 3s, 3p and Cd 5s, 5p states. An intense peak in the total DOS in the range -5.4 to -2 eV originate from the strong hybridization of Ti 3d and C 2p states indicating the covalent Ti-C bonds in Ti$_2$CdC. At the Fermi level, the DOS mainly arises from the Ti 3d states. The calculated DOSs at the Fermi level $N(E_F)$ is 5.63 states per unit cell per eV. Our results are consistent with the value of 5.595 states per unit cell per eV calculated by Bai et al. [14]. The conduction properties of Ti$_2$CdC result due to the Ti 3d contribution. This is consistent with the previous reports published on other MAX phases [32, 35]. Almost similar features are found for Ti$_2$CdN phase, though the lowest-lying valence bands disappear in Ti$_2$CdC when C is substituted by N. However, the DOSs at the $E_F$ decrease from 5.63 to 4.57 states per unit cell per eV as C is replaced with N. This is consistent with the results observed in Ti$_2$AlX [36, 37] and Ti$_2$InX [38], but differs with the calculations for Ti$_2$AlX by Du et al. [27] and for Ti$_2$InX by Benayad et al. [39]. From the calculated band structure, the overall bonding character in two MAX phases may be described as a mixture of metallic, covalent and, due to the difference in the electronegativity between the constituting elements, ionic.

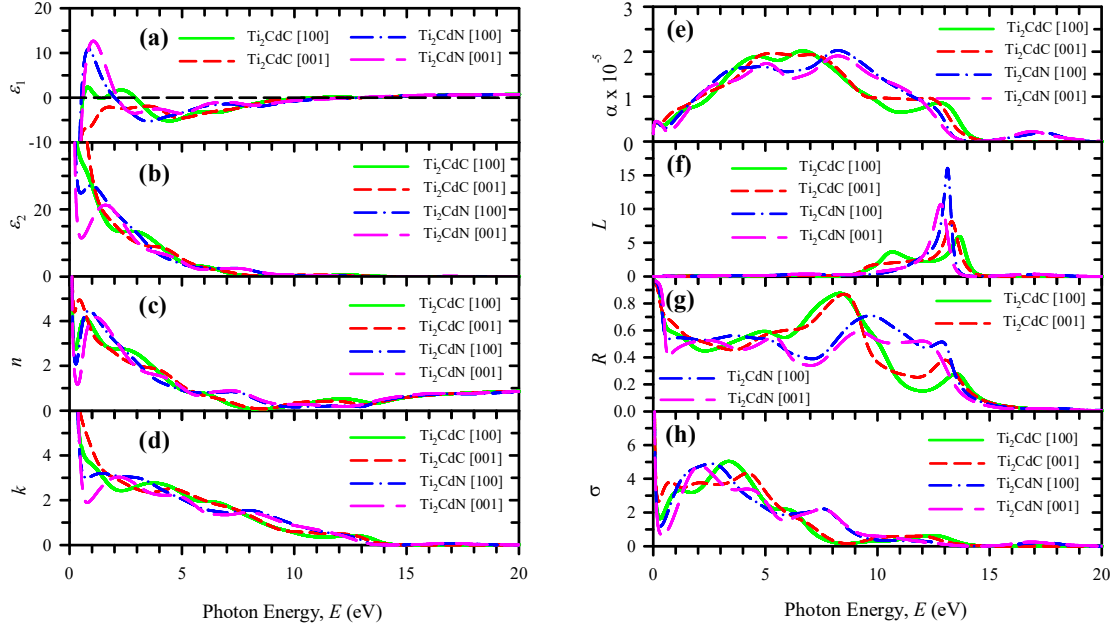

**Fig. 4**. Optical parameters of Ti$_2$CdC and Ti$_2$CdN for two different polarizations of incident radiation. (a) Real part of the dielectric function, (b) the imaginary part of the dielectric function, (c) the real part of the refractive index, (d) the imaginary part of the refractive index, (e) the absorption coefficient, (f) the energy-loss function, (g) the reflectivity and (h) the real part of the optical conductivity.

### *3.4. Optical properties*

The calculated optical parameters (dielectric function, the refractive index, the absorption spectrum, the energy loss function, the reflectivity and the optical conductivity) of the two MAX compounds for the incident photon energies up to 20 eV for the two polarization vectors along [100] and [001] are shown in Fig. 4. The dielectric function characterizes how a material responds to the incident electromagnetic wave. The real and imaginary parts of the dielectric function are plotted in Figs. 4(a) and 4(b), respectively. The real part of the dielectric function $\varepsilon_1(\omega)$ of Ti$_2$CdC for the polarization along [100] direction exhibit two peaks at around 0.8 and 2.3 eV, whereas for [001] direction no peak is found. In case of Ti$_2$CdN, a peak is observed at 0.88 eV for the direction [100] and at 1.1 eV for the direction [001]. Both $\varepsilon_1(\omega)$ and $\varepsilon_2(\omega)$ show metallic characteristics in agreement with the general feature of other MAX phases [26, 32, 33, 40]. In high energy region the real part of the dielectric function tends to be unity and the imaginary part reaches nearly zero. It implies that in this region these materials become almost transparent with very little absorption. From Figs 4(a) and 4(b), it is evident that both the phases are optically anisotropic.

The real and imaginary parts of the refractive indices are presented in Figs. 4(c) and 4(d). The real part of the refractive index indicates the phase velocity, while the imaginary part, often termed as extinction coefficient, indicates the amount of absorption loss when the electromagnetic wave propagates through the material. In spite of some variation in heights and positions of peaks, the overall features of real and imaginary parts of the refractive indices for the two MAX phases are qualitatively same in the entire energy range.

Fig. 4(e) exhibits the absorption spectra. The absorption coefficient determines how far light of a particular wavelength (energy) can penetrate into a material before it gets absorbed. It also provides information about the optimum solar energy conversion efficiency. The absorption spectra rise sharply below 3 eV. The spectra show two peaks between 3.2 and 8.5 eV and then continue to decrease up to 13.5 eV. These peaks arise due to transition from Cd/X *p* to Ti *d* states. The optical anisotropy found in the absorption spectra of the two nanolaminates is very small. The peak-positions shift when C is replaced by N in Ti$_2$CdC. Both the phases show large absorption coefficient in the 3-10 eV region.

The energy loss spectra are presented in Fig. 4(f). The energy loss function of a material is an important parameter in the dielectric formalism which is useful for understanding the screened excitation spectra, especially the collective excitations produced by the swift charges inside a solid. The highest peak of the energy loss spectrum appears at a particular incident light frequency (energy) known as the bulk plasma frequency. For the polarization direction [100], the plasma frequency of Ti$_2$CdC and Ti$_2$CdN are found to be 13.6 and 13.1 eV, respectively. On the other hand, the corresponding values for the two phases for the incident light direction [001] are seen to be 13.3 and 12.8 eV, respectively.

The reflectivity is the ratio of the energy of a wave reflected from a surface to the energy of the wave incident on the surface. At around 5.2 eV, a gradual increase starts for Ti$_2$CdC but a rapid decrease occurs in reflectivity spectra of Ti$_2$CdN. An intense peak observed at around 8.5 eV in the reflectivity spectra of Ti$_2$CdC is shifted at around 9.8 eV when C is replaced with N. In the moderate-infrared region, the reflectivity spectra of both increase drastically and rise to reach a maximum value of 0.98. With almost a constant value in the region of 1.8-5.1 eV, the reflectivity spectra of the two phases for both the polarization directions show no significant change which is advantageous in application as a coating material [26]. Moreover, the strong edge and color are absent in the reflectance of these MAX compounds. One can thus conclude that the two MAX phases Ti$_2$CdC and Ti$_2$CdN show the nonselective characteristic that makes them suitable candidates for reducing solar heating. Compared to Ti$_2$CdN, the reflectance of Ti$_2$CdC is always smaller for both the incident light directions within the energy range of 2.0-3.8 eV, while it is large inside the 4.1-9.2 eV region. Therefore, Ti$_2$CdN might be a better candidate material for coating to avoid solar heating in the lower energy region. However, in the higher energy region Ti$_2$CdC might be more suitable.

The real part of the optical conductivity of Ti$_2$CdC and Ti$_2$CdN are compared in Fig. 4(h). The optical conductivity is expected to be a good gauge of the photoconductivity [41]. The optical conductivity shows a sharp dip from 0 to 0.67 eV and then increases to reach the maximum value of ~ 4.3-5.0 in the energy range 2.0-4.3 eV. Consequently, the two Cd-containing MAX phases should be more electrically conductive in the incident photon energy range of 2.0 to 4.3 eV.

## 4. Conclusion

First-principles calculations have been performed to investigate the structural, elastic, electronic and optical properties of the Cd-containing MAX phases Ti$_2$CdC and Ti$_2$CdN. Investigation of phase stability via enthalpy calculations shows that the predicted Ti$_2$CdN should be stable against decomposition into binary constituents. Our results show that the substitution of C by N in Ti$_2$CdC affects all properties such as structural, elastic, electronic, etc. The replacement of C with N increases all the elastic constants and moduli of Ti$_2$CdC. The predicted phase Ti$_2$CdN appears to be more important than the already synthesized Ti$_2$CdC in many engineering applications since it is found to be stiffer than Ti$_2$CdC. The elastic anisotropy of Ti$_2$CdC is higher compared to Ti$_2$CdN. Both the MAX phases are found to be brittle in nature and Ti$_2$CdN is more brittle than Ti$_2$CdC. The chemical bonding in two nanolaminates is seen to be a combination of covalent, ionic and metallic nature. Ti$_2$CdN is less conducting than Ti$_2$CdC and consequently the ceramic properties of it are more suitable for high temperature applications. The MAX phases Ti$_2$CdC and Ti$_2$CdN are optically anisotropic. The ability of Ti$_2$CdN to reduce solar heating is stronger than Ti$_2$CdC in the visible region.

In conclusion, we hope that these theoretical findings will stimulate experimental effort to synthesize the predicted Ti$_2$CdN Max phase, which has several properties of interest for a wide range of applications.